\documentclass[aps, prc, reprint, amsmath, amssymb, superscriptaddress, floatfix, nofootinbib]{revtex4-1} 

\usepackage[pdftex]{graphicx}
\DeclareGraphicsExtensions{.jpg,.png,.pdf}

\usepackage[utf8]{inputenc}
\usepackage{hyperref}
\usepackage{amsmath}
\usepackage{amssymb}
\usepackage{amsfonts}
\usepackage{tabularx}
\usepackage{booktabs}
\usepackage{graphicx}
\usepackage{subfigure}
\usepackage{wrapfig}
\usepackage{float}
\usepackage{color}
\usepackage{multirow}
\usepackage[perpage,symbol]{footmisc}
\usepackage{verbatim}
\usepackage{dcolumn}
\usepackage{bm}
\usepackage[inline]{enumitem}
\definecolor{theblue}{RGB}{0,50,230}
\hypersetup{
  colorlinks=true,
  linkcolor=theblue,
  citecolor=theblue,
  urlcolor=theblue
}
\newcommand{\pt}{\ensuremath{p}_{\rm T}}
\newcommand{\raa}{\ensuremath{R}_{\rm AA}}
\newcommand{\vtwo}{\ensuremath{v}_{\rm 2}}

\newcommand{\snn}{\sqrt{s_{\rm NN}}}
\newcommand{\s}{\sqrt{s}}

\usepackage{lineno}
\begin{document}

\title{Relativistic Langevin dynamics: charm versus beauty}

\author{Shuang~Li}                                                                                                                                                           
\email{lish@ctgu.edu.cn}
\affiliation{%
College of Science, China Three Gorges University, Yichang 443002, China\\
}%
\affiliation{%
Key Laboratory of Quark and Lepton Physics (MOE), Central China Normal University, Wuhan 430079, China\\
}%

\author{Wei Xiong}%
\email{xiongw@ctgu.edu.cn}
\affiliation{%
College of Science, China Three Gorges University, Yichang 443002, China\\
}%
\author{Renzhuo Wan}%
\email{wanrz@wtu.edu.cn}
\affiliation{%
Nano Optical Material and Storage Device Research Center, School of Electronic and Electrical Engineering,\\
Wuhan Textile University, Wuhan 430200, China\\
}%

\date{\today}

\begin{abstract}
The production of heavy quarks (charm and beauty) provides unique insights into the transport properties
of the Quark-Gluon Plasma (QGP) in heavy-ion collisions.
Experimentally, the nuclear modification factor $\raa$ and the azimuthal anisotropy coefficient $\vtwo$
of heavy-flavor mesons are powerful observables to study the medium-related effects,
such as energy loss and collectivity, on the heavy quark propagation through the QGP evolution.
The latest measurements of the prompt and non-prompt open heavy-flavor hadrons
allow a systematic comparison of the transport behaviors probed by charm and beauty quarks.
In this work we make such an attempt utilizing our recently developed framework.
By performing a quantitative investigation of $\raa$ and $\vtwo$,
it is found that both charm and beauty quarks are efficient probes to capture the dynamical features of QGP,
in particular the resulting mass hierarchy for the energy loss and azimuthal anisotropy,
which are well inherited by the various $D/B$-meson species.
Moreover, our calculations can describe simultaneously
$\raa$ and $\vtwo$ data for the prompt and non-prompt $D^{0}$ mesons
in central ($0-10\%$) and semi-central ($30-50\%$) Pb--Pb collisions at $\snn=5.02~{\rm TeV}$.
The predictions for $B$-meson observables for upcoming experimental tests are also made down to the low momentum region.
\end{abstract}


\maketitle

\section{Introduction}\label{sec:Intro}
Experimental measurements of open heavy-flavor productions (having charm or beauty quarks among these valence quarks)
are regarded as efficient probes to extract the properties of the deconfined nuclear matter
known as Quark-Gluon Plasma (QGP), in ultrarelativistic heavy-ion collisions~\cite{Rapp:2018qla,CaoCoefficient18,Dong:2019byy,Zhou:2014kka,Tang:2014tga,Andronic:2015wma}.
This is because, due to large mass, heavy quarks (HQ) are very hard to be thermally produced in QGP
and are dominantly produced from the initial hard scatterings.
Subsequently, they propagate through the QGP
and experience its full evolution by interacting
elastically and inelastically with the its constituents, e.g. the (anti-)light quarks and gluons~\cite{PBGNPA19}.
The resulting effects on HQ, such as the energy loss and collectivity, have been found to be particularly informative for
unraveling the dynamical features of QGP~\cite{Gyulassy05,Shuryak05,Muller12}.

While traversing the QGP medium,
HQ suffer multiple scatterings and behaves a Brownian motion,
which can be quantified by a Boltzmann Transport Equation (BTE~\cite{Benjamin88}).
It is argued~\cite{RalfSummary16, ZhuangHFSummary20} that, for massive quarks at moderate medium temperatures,
the typical momentum transfers are small,
therefore, BTE reduces to a Langevin Transport Equation (LTE).
In the framework of the stochastic Langevin approach,
the knowledge of three transport coefficients in momentum space,
representing the drag, transverse and longitudinal momentum broadening behaviors,
is required to trace the HQ propagation in the medium.
With the help of the dissipation-fluctuation relation,
all the mentioned coefficients can be quantified by the intrinsic medium transport parameter, $2\pi TD_{s}$,
consequently, the HQ-medium interactions are conveniently encoded in this single parameter.
Many works are developed from both the Boltzmann~\cite{BAMPS10, DjordjevicPLB14, LBTPRC16, PHSDPRC16, XWJPLB20}
and Langevin dynamics~\cite{Das10, HFModelHee13, CaoPRC15, MCATHQPRC16, AIS19, CAGP19, SSCNewCoal19}
to investigate its effect on the suppression and collectivity of open charm hadrons.
It was realized~\cite{JFLPRL09, Das15, JFLCPL15, CTGUHybrid1} that the simultaneous description of $\raa$ and $\vtwo$
at low and intermediate transverse momentum is sensitive to the temperature dependence of $2\pi TD_{s}(T)$.
To nail down its temperature dependence
and to improve the description of the open charm hadron measurements,
a linear ansatz was adopted in our previous work
within the region accessible through RHIC and LHC collision experiments,
and the adjustable parameters were optimized, via a $\chi^2$ analysis,
by comparing the comprehensive sets of $D$-meson data at LHC energy
with the calculations from our recently developed framework LGR (Langevin-transport with Gluon Radiation)~\cite{CTGUHybrid4}.
It was argued~\cite{CTGUHybrid4} that a relatively strong increase of $2\pi TD_{s}$ from crossover temperature $T_c$
toward high temperature was preferred by the available measurements.

Recently, ALICE Collaboration reports the first comparisons of $\raa$ between non-prompt and prompt $D^{0}$ mesons down to low transverse momentum,
where will be more informative for the understanding of $2\pi TD_{s}(T)$.
Thus, in this work, we plan to study the medium-induced effects on the $\raa$ and $\vtwo$ between charm and beauty by using the LGR model.
The relevant comparisons with data and the other modeling framework,
allow us to check again the stability of the obtained $2\pi TD_{s}(T)$,
as well as the mass hierarchy for the HQ energy loss and azimuthal anisotropy.

The paper is organized as follows:
Section~\ref{sec:Method} is dedicated to discussing in detail the relevance of the LGR modeling framework.
In Sec.~\ref{sec:Result}, we systematically compare the energy loss and coalescence mechanism between charm and beauty quarks in the LGR approach.
Also, the modeling results of $\raa$ and $\vtwo$ for prompt and non-prompt productions
are compared with data and the CUJET3 model.
Finally, we summarize this study in Sec.~\ref{sec:Summary}.

\section{Methodology}\label{sec:Method}
The Langevin-transport with Gluon Radiation (LGR) is expressed as
\begin{equation}
\begin{aligned}\label{eq:LTE_ColRad}
&\frac{d\vec{x}}{dt}=\frac{\vec{p}}{E}
& \\
&\frac{d\vec{p}}{dt}=\vec{F}_{\rm Drag} + \vec{F}_{\rm Thermal} + \vec{F}_{\rm Recoil}.
\end{aligned}
\end{equation}
The deterministic drag force received from the surrounding medium constituents can be described by
\begin{equation}\label{eq:DragForce}
\vec{F}_{\rm Drag}=-\eta_{\rm D} \cdot \vec{p},
\end{equation}
where, $\eta_{\rm D}$ is the drag coefficient.
The stochastic thermal force reads 

\begin{equation}\label{eq:ThermalForce}
F^{i}_{\rm Thermal}=\frac{1}{\sqrt{dt}}C^{ij}\rho^{j}
\end{equation}
with $i,j=1,2,3$. The Gaussian noise $\rho^{j}$ follows a normal distribution $ P(\vec{\rho})=(\frac{1}{2\pi})^{3/2} exp\{-\frac{\vec{\rho}^{\;2}}{2\;}\}$,
resulting in the uncorrelated random momentum kicks between two different time scales.
The momentum argument of the covariance matrix, $C^{ij}$,
is given by the longitudinal ($\kappa_{\parallel}$) and transverse momentum diffusion coefficients ($\kappa_{\perp}$)~\cite{POWLANG09}
\begin{equation}
\begin{aligned}\label{eq:LTEtensor}
C^{\rm ij} &\equiv \sqrt{\kappa_{\parallel}}\frac{p^{i}p^{j}}{\vec{p}^{\;2}} + \sqrt{\kappa_{\perp}}(\delta^{ij}-\frac{p^{i}p^{j}}{\vec{p}^{\;2}}) =\sqrt{\kappa}\delta^{ij}.
\end{aligned}
\end{equation}
Note that the momentum dependence of the diffusion coefficient
is assumed to be isotropic ($\kappa_{\parallel} = \kappa_{\perp} \equiv \kappa$~\cite{CTGUHybrid1}) to obtain the second step in Eq.~\ref{eq:LTEtensor}.
According to the dissipation-fluctuation relation in the non-relativistic approximation,
the drag coefficient can be represented in terms of the momentum diffusion coefficient
\begin{equation}\label{eq:PostPoint2}
\eta_{\rm D}=\frac{\kappa}{2TE},
\end{equation}
with the post-point discretization scheme.
The total recoil force induced by the emitted gluons is expressed as 
\begin{equation}\label{eq:RecoilForce}
\vec{F}_{\rm Recoil}=-\sum^{N_{G}}_{j=1}\frac{d\vec{p}^{\;j}_{\rm G}}{dt},
\end{equation}
and the single medium-induced emission rate is predicted with the Higher-Twist approach~\cite{HTPRL04}
\begin{equation}\label{eq:HigherTwist}
\frac{dN_{G}}{dz dk_{\perp}^{2} dt}=\frac{2\alpha_{s}C_{\rm A}P(z) \hat{q}_{\rm q}}{\pi k_{\perp}^{4}}
\biggr[\frac{k_{\perp}^{2}}{k_{\perp}^{2}+(zm_{\rm Q})^{2}}\biggr]^{4} sin^{2}\biggr( \frac{t-t_{0}}{2\tau_{f}} \biggr),
\end{equation}
where, $z=k/E$ denotes the fraction of energy carried away by the emitted gluon,
and $P(z)=(z^{2}-2z+2)/z$ represents the quark splitting function;
$\alpha_{s}(k_{\perp})=\frac{4\pi}{11N_{c}/3-2N_{f}/3}({\rm ln}\frac{k_{\perp}^{2}}{\Lambda^{2}})^{-1}$
is the strong coupling constant of QCD at leading order approximation;
$\tau_{f}=2z(1-z)E/[k_{\perp}^{2}+(zm_{\rm Q})^{2}]$ is the gluon formation time;
$\hat{q}_{\rm q}$ is the quark jet transport coefficient,
which can be approximated by $\hat{q}_{\rm q}\approx 2\kappa_{\perp}$ in the high energy region.
On the right-hand side of Eq.~\ref{eq:HigherTwist},
the second term in the bracket represents the heavy quark mass effect, the so-called dead-cone effect~\cite{DeadConePLB01}.
The third term reveals the suppression behavior of the radiated gluon spectra,
which is induced by the coherent multiple inelastic scatterings during the time-scale $\tau_{f}$,
namely, the Landau-Pomeranchuck-Migdal (LPM) effect~\cite{LPMNPB94}.

With the help of the (scaled) spacial diffusion coefficient $2\pi TD_{s}(T)$,
the drag and the momentum diffusion coefficients (Eq.~\ref{eq:PostPoint2}) can be represented as~\cite{CTGUHybrid1, CTGUHybrid3}
\begin{equation}
\begin{aligned}\label{eq:LTECoef}
&\eta_{\rm D}(\vec{p},T)=\frac{1}{2\pi TD_{s}} \cdot \frac{2\pi T^{2}}{E} \\
&\kappa(T)=\frac{1}{2\pi TD_{s}} \cdot {4\pi T^{3}}.
\end{aligned}
\end{equation}
We can see that there is only one parameter ($2\pi TD_{s}$) left to
quantify the drag force (Eq.~\ref{eq:DragForce}), thermal random force (Eq.~\ref{eq:ThermalForce})
and the recoil force (Eq.~\ref{eq:RecoilForce} and~\ref{eq:HigherTwist})
in the Langevin approach (Eq.~\ref{eq:LTE_ColRad}).
Therefore, the interactions between the HQ and the incident medium constituents are conveniently encoded into
the temperature-dependence of $2\pi TD_{s}(T)$.
In our previous work, $2\pi TD_{s}(T)$ for charm quark is approximated by~\cite{CTGUHybrid4}
\begin{equation}\label{eq:Assum}
2\pi TD_{s}(T) \approx \alpha\frac{T}{T_{c}} + \beta
\end{equation}                                                                                                                                                               
within a wide range $T_{c}<T\lesssim3T_{c}$.
Moreover, the free parameters, accounting for the slope $\alpha=6.5$ and intercept $\beta=-5.5$,
are optimized by using a data-driven $\chi^{2}$ analysis~\cite{CTGUHybrid4}.
Concerning $2\pi TD_{s}(T)$ for beauty quark,
it is estimated by scaling the charm quark result with a constant factor,
$2\pi TD_{s}(beauty) \approx 0.85\times 2\pi TD_{s}(charm)$, 
since the ratio of $2\pi TD_{s}$ between charm and beauty presents a weak $T$-dependence and
varies within $\sim0.8-0.9$ in the range $T_{c}<T<4T_{c}$~\cite{CTGUHybrid3,POWLANGEPJC11,DasPRD16}.

During the numerical implementation of the Langevin transport equation (Eq.~\ref{eq:LTE_ColRad}),
first, following our previuos work~\cite{CTGUHybrid1,CTGUHybrid2,CTGUHybrid3},
the initial momentum spectra of heavy quark pairs are
given by the FONLL pQCD calculations~\cite{FONLL98, FONLL01, FONLL12}.
Then, we need the space-time evolution of the temperature and the fireball velocity field.
It is simulated in terms of a 3+1 dimensional relativistic viscous hydrodynamics based on the HLLE algorithm~{\cite{vhlle}},
with the local thermalization started at $\tau_{0}=0.6~{\rm fm}/{\it c}$,
the shear viscosity over entropy ratio $\frac{\eta}{s}=\frac{1}{4\pi}$
and the critical temperature $T_{c}=165~{\rm MeV}$ in high energy heavy-ion collisions.
Its initial entropy density is modeled by a Glauber approach~\cite{iEBE},
while the Equation of State (EoS) given by the lattice QCD predictions~\cite{EoSLaine06}.
Finally, a ``dual" approach, including fragmentation and heavy-light coalescence mechanisms,
is utilized to describe the HQ hadroniztion process when the local temperature below $T_{c}$.
The Braaten approach~\cite{FragBraaten93} is employed
to complete the fragmentations for both charm and beauty quarks.
Within the instantaneous coalescence approach,
there is the probability for a heavy quark ($Q$) to form the relevant heavy-flavor meson ($M$),
by hadronizing via coalescence with a thermal parton ($\bar{q}$) from the fireball.
The resulting momentum distributions reads
\begin{equation}
\begin{aligned}\label{eq:MesonCoal}
\frac{dN_{\rm M}}{d^{3}\vec{p}_{\rm M}}=g_{\rm M} \int d^{6}\xi_{\rm Q} d^{6}\xi_{\rm\bar{q}} f_{\rm Q} f_{\rm\bar{q}}
\sum_{n=0}^{1} {\overline W}_{\rm M}^{(n)} \delta^{(3)}(\vec{p}_{\rm Q}+\vec{p}_{\rm\bar{q}}-\vec{p}_{\rm M})
\end{aligned}
\end{equation}
where, $g_{\rm M}$ indicates the spin-color degeneracy factor;
$d^{6}\xi_{i}=d^{3}\vec{x}_{i}d^{3}\vec{p}_{i}$ denotes the phase-space volume for parton $i=Q,\bar{q}$;
$f_{i}(\vec{x}_{i},\vec{p}_{i})$ is the phase-space distributions;
${\overline W}_{\rm M}^{(n)}$ represents the heavy-light coalescence probability for $Q\bar{q}$ combination
to form the heavy-flavor meson in the $n^{th}$ excited state,
and it is defined as the overlap integral of the Wigner functions for the meson and $Q\bar{q}$ pair~\cite{NewCoal16}
\begin{equation}\label{eq:InteWigner2}
{\overline W}_{\rm M}^{\rm (n)}(\vec{y}_{\rm M},\vec{k}_{\rm M})=\frac{\lambda^{n}}{n!} e^{-\lambda}; \quad
\lambda=\frac{1}{2}\biggr(\frac{\vec{y}_{\rm M}^{\;2}}{\sigma_{\rm M}^{2}}+\sigma_{\rm M}^{2}\vec{k}_{\rm M}^{\;2}\biggr),
\end{equation}
where, the relative coordinate and momentum,
\begin{equation}
\begin{aligned}\label{eq:RelYK}
&\vec{y}_{\rm M} \equiv \vec{x}_{\rm Q}-\vec{x}_{\rm\bar{q}}
& \\
&\vec{k}_{\rm M} \equiv (m_{\rm\bar{q}}\vec{p}_{\rm Q}-m_{\rm Q}\vec{p}_{\rm\bar{q}})/(m_{\rm Q}+m_{\rm\bar{q}}),
\end{aligned}
\end{equation}
are defined in the center-of-mass frame of $Q\bar{q}$ pair;
the relevant width parameter $\sigma_{\rm M}$ is given by
\begin{eqnarray}\label{eq:SigM}
\sigma_{\rm M}^{2} = K \frac{(m_{\rm Q}+m_{\rm\bar{q}})^{2}}{e_{\rm Q}m_{\rm\bar{q}}^{2}
+ e_{\rm\bar{q}}m_{\rm Q}^{2}} \langle r_{\rm M}^{2} \rangle,
\end{eqnarray}
with $K=2/3$ ($K=2/5$) for the ground state $n=0$ ($1^{st}$ excited state $n=1$);
$\langle r_{\rm M}^{2} \rangle$ denotes the mean-square charge radius of a given species of $D/B$-meson,
which can be estimated according to the light-front quark model~\cite{HwangEPJC02}.                                                                                          
See Ref.~\cite{CTGUHybrid1,CTGUHybrid3} for more details.

\section{Results and discussions}\label{sec:Result}
\begin{figure}[!htbp]
\centering
\setlength{\abovecaptionskip}{-0.1mm}
\includegraphics[width=.40\textwidth]{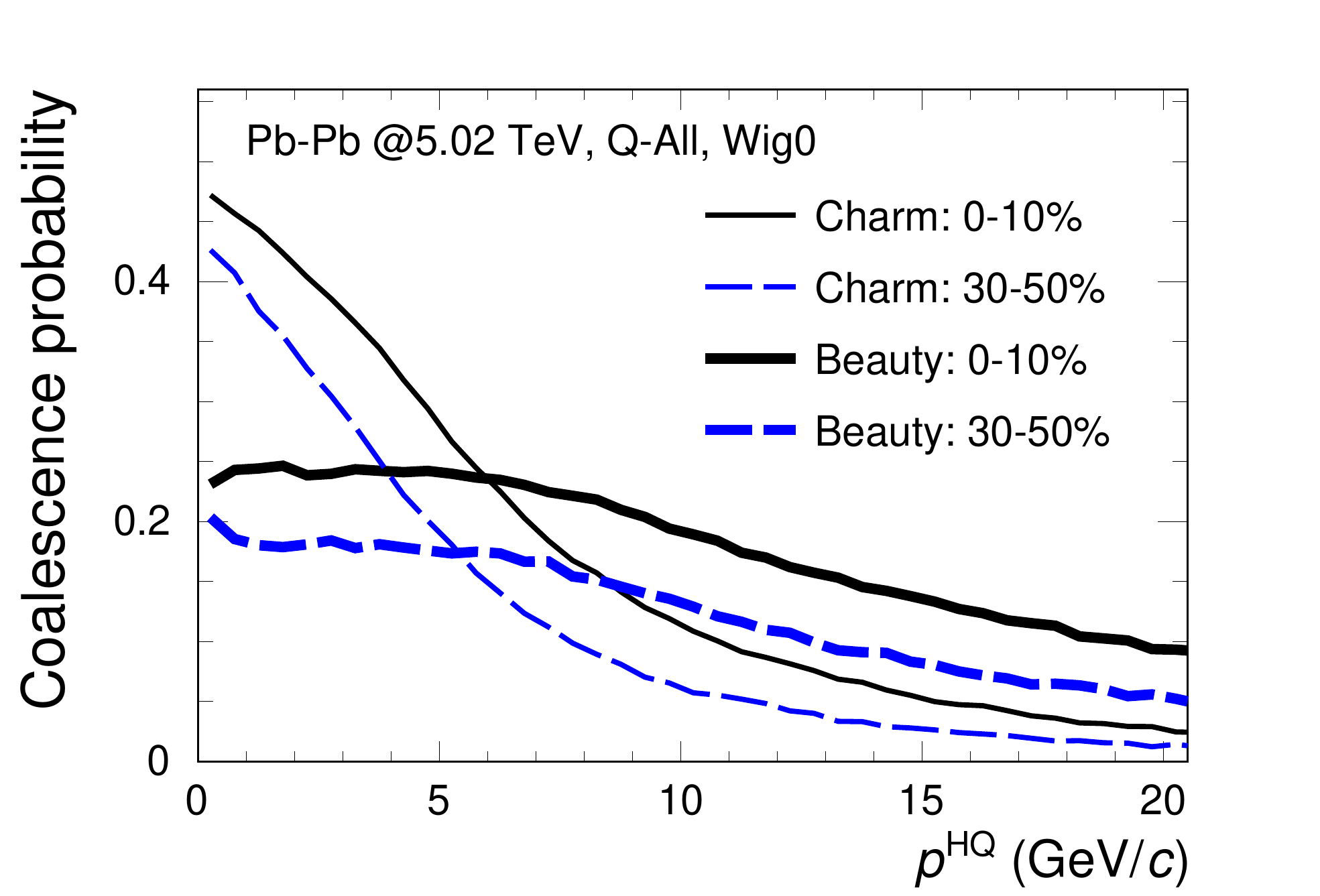}
\caption{Comparison of the coalescence probability,
for $c\bar{q}$ (thin curves) and $b\bar{q}$ combinations (thick curves) in central $0-10\%$ (solid black curves) and                                                         
semi-central $30-50\%$ (long dashed blue curves) Pb--Pb collisions at $\snn=5.02~{\rm TeV}$,
to form the $D$ ($c\bar{q}$) and $B$ mesons ($b\bar{q}$) at the ground state.}
\label{fig:CoalProb_vsP_CB_5020}
\end{figure}

In Fig.~\ref{fig:CoalProb_vsP_CB_5020}, the total coalescence probability to form
the various species of $D$ ($c\bar{q}$) and $B$ mesons ($b\bar{q}$) with the ground state,
is calculated in central ($0-10\%$) and semi-central ($30-50\%$) Pb--Pb collisions at $\snn=5.02~{\rm TeV}$.
It is interesting to see that:
\begin{enumerate}
\item[(1)] most of the results have maximum probability at $p^{\rm HQ}\sim0$,
and it decreases towards high $p^{\rm HQ}$, due to the difficulty to find a coalescence partner in this region;
\item[(2)] results are systematically larger in more central collisions,
since the relevant partner density is larger in this region, resulting in a larger probability to form heavy-light combinations;
\item[(3)] within a desired centrality class,
${\overline W}_{\rm M}^{\rm (0)}(c\bar{q})$ is systematically larger (smaller) than
${\overline W}_{\rm M}^{\rm (0)}(b\bar{q})$ in the range $p^{\rm HQ} < 4-5~{\rm GeV/{\it c}}$ ($p^{\rm HQ} > 4-5~{\rm GeV/{\it c}}$).
This behavior may be induced by the fact that,
I) ${\overline W}_{\rm M}^{\rm (0)}\sim exp\{ -\frac{1}{2}\sigma_{\rm M}^{2}\vec{k}_{\rm M}^{\;2}\}$:
${\overline W}_{\rm M}^{\rm (0)}$ is mostly affected
by $|\vec{k}_{\rm M}|$ rather than $|\vec{y}_{\rm M}|$, since the coalescence partners are sampled nearby a HQ is spacial space~\cite{CTGUHybrid1};
II) $\sigma^{2}_{\rm M}\sim 2\langle r^{2}_{\rm M} \rangle$ is very similar
for a given $c\bar{q}$ and $b\bar{q}$ combinations, e.g. $\langle r^{2}_{D^{+}_{s}(c\bar{s})} \rangle = 0.124~{\rm fm^{2}}$
and $\langle r^{2}_{\bar{B}^{0}_{s}(b\bar{s})} \rangle = 0.119~{\rm fm^{2}}$~\cite{HwangEPJC02};
III) $|\vec{k}_{\rm M}|$ behaves\footnote[5]{Note that in Eq.~\ref{eq:EstRealMom}
$\vec{p}_{\rm Q}$ and $\vec{p}_{\rm \bar{q}}$ denote, respectively, the 3-momentum of HQ and its coalelcence partner
in the laboratory frame, while in Eq.~\ref{eq:RelYK} they are the similar ones but defined in the center-of-mass frame of $Q\bar{q}$ system.}
\begin{eqnarray}\label{eq:EstRealMom}
|\vec{k}_{\rm M}| \sim \frac{s-m^{2}_{\rm Q}}{2\sqrt{s}}
&&\sim \left\{ \begin{array}{ll}
\frac{E_{\bar{q}}}{1+\frac{2E_{\bar{q}}}{m_{\rm Q}}} & \textrm{($p_{\rm Q} \ll m_{\rm Q}$)} \\
\\
\frac{p_{\rm Q}E_{\bar{q}}}{\sqrt{m^{2}_{\rm Q}+2E_{\bar{q}}p_{\rm Q}}} & \textrm{($p_{\rm Q} \gg m_{\rm Q}$)} 
\end{array} \right.
\end{eqnarray}
by assuming $\vec{p}_{\rm Q} \cdot \vec{p}_{\rm \bar{q}} \sim 0$ for simplicity.
Therefore, for a given HQ ($p_{\rm Q}$) and partner ($E_{\bar{q}}$),
$|\vec{k}_{\rm M}|$ is smaller (larger) for charm at low (high) $p_{\rm Q}$,
leading to a larger (smaller) ${\overline W}_{\rm M}^{\rm (0)}$ as compared to that for beauty quarks.
\end{enumerate}

\begin{figure}[!htbp]
\centering
\setlength{\abovecaptionskip}{-0.1mm}
\includegraphics[width=.40\textwidth]{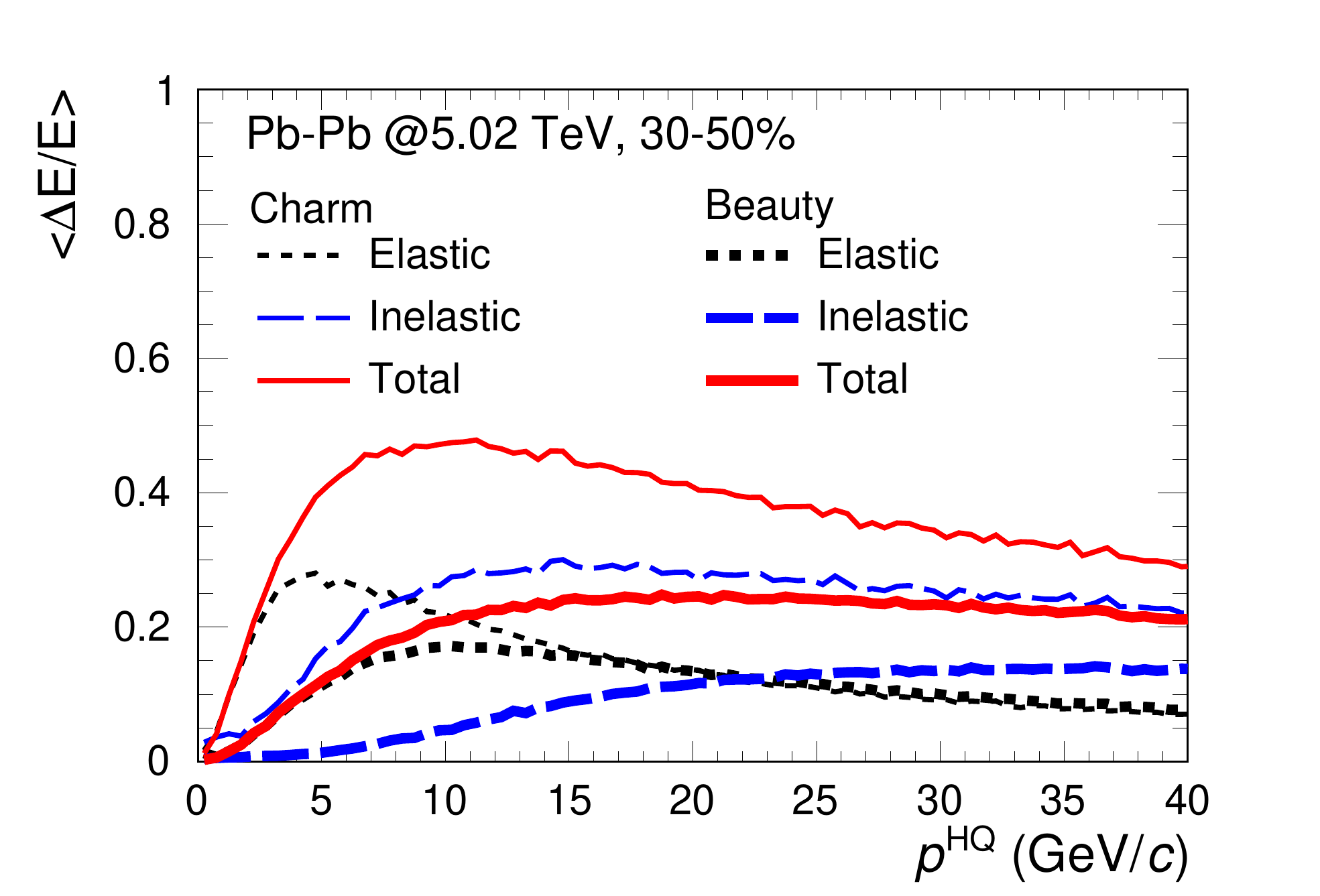}
\caption{Comparison of the average in-medium energy loss fraction,
for charm (thin curves) and beauty quarks (thick curves),
displaying separately the contributions from collisional (dashed black curves) and radiative mechanisms (long dashed blue curves).
The combined results (solid red curves) are shown as well for comparison.}
\label{fig:DeltaE_CB_5020}
\end{figure}
Figure~\ref{fig:DeltaE_CB_5020} presents the average in-medium energy loss fraction of charm (thin curves)
and beauty quark (thick curves) as a function of its initial momentum,
displaying separately the contributions from elastic (dashed black curves) and inelastic scatterings (long dashed blue curves).
It is found that:
\begin{enumerate}
\item[(1)] collisional component for charm (beauty) quark
shows a increasing behavior in the range $p^{\rm HQ}\lesssim4~{\rm GeV/\it{c}}$ ($p^{\rm HQ}\lesssim10~{\rm GeV/\it{c}}$),
followed by a decreasing trend at higher $p$.
This is because, I) the elastic scatterings are dominated by the drag (Eq.~\ref{eq:DragForce})
rather than the diffusion term (Eq.~\ref{eq:ThermalForce}),
since the initial HQ spectrum is much more harder that of medium partons;
II) the drag force acted on the HQ behaves as $F_{\rm Drag}\propto p/E$ (Eq.~\ref{eq:DragForce} and \ref{eq:LTECoef}) at a given temperature,
and the induced energy loss fraction follows $\Delta{E}/E\propto p/(p^{2}+m^{2}_{\rm Q})$;
thus, $\Delta{E}/E$ for charm quark ($m_{\rm c}=1.5~{\rm GeV}$) in the low momentum region $p^{\rm HQ}\lesssim2m_{\rm Q}$,
is expected to be larger than that for beauty quark ($m_{\rm b}=4.75~{\rm GeV}$),
while in the very large momentum region $p^{\rm HQ}\gg m_{\rm Q}$, the mass effect can be neglected,
resulting in a similar behavior between them;
they are consistent with the observations shown above;
\item[(2)] radiative energy loss fraction can be calculated via
\begin{equation}
\begin{aligned}\label{eq:ELoss_HT}
<\frac{\Delta{E}}{E}>=\int_{0}^{t}dt^{\prime} \int_{\frac{\omega_{0}}{E}}^{1}dz \int_{0}^{(zE)^{2}}dk_{\perp}^{2}z\frac{dN_{G}}{dz dk_{\perp}^{2} dt^{\prime}},
\end{aligned}
\end{equation}
where, a lower cut-off $\omega_{0}=\pi T$ is imposed on the gluon energy~\cite{CaoPRC15}.
Equation~\ref{eq:ELoss_HT} together with Eq.~\ref{eq:HigherTwist} is difficult to be calculated analytically
within a realistic hydrodynamic medium.
However, as discussed in Eq.~\ref{eq:HigherTwist},
due to the mass effect ($m_{\rm b}>m_{\rm c}$), charm quark has a larger probability of single bremsstrahlung
with a given temperature and time-interval,
resulting in stronger radiative energy loss effect when comparing with beauty;
the calculations for charm and beauty are expected to agree in asymptotically large momentum region,
which are consistent with the results as shown in Fig.~\ref{fig:DeltaE_CB_5020};
\item[(3)] collisional energy loss is significant at low momentum,
while radiative energy loss is the dominant mechanism at high momentum.
It should be noticed that similar behavior can be found for the results as a function of transverse momentum.
\end{enumerate}

\begin{figure}[!htbp]
\centering
\setlength{\abovecaptionskip}{-0.1mm}
\includegraphics[width=.40\textwidth]{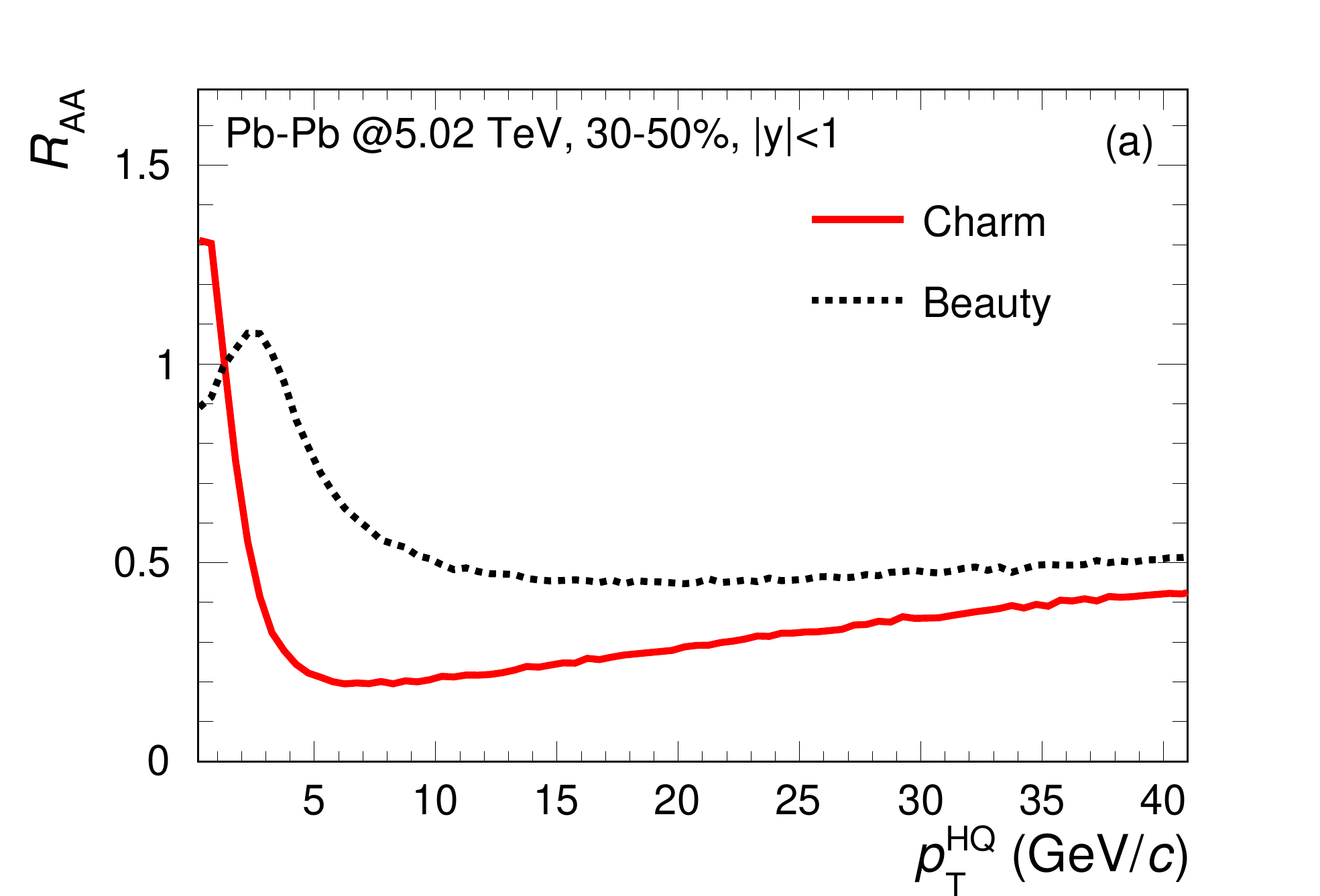}
\includegraphics[width=.40\textwidth]{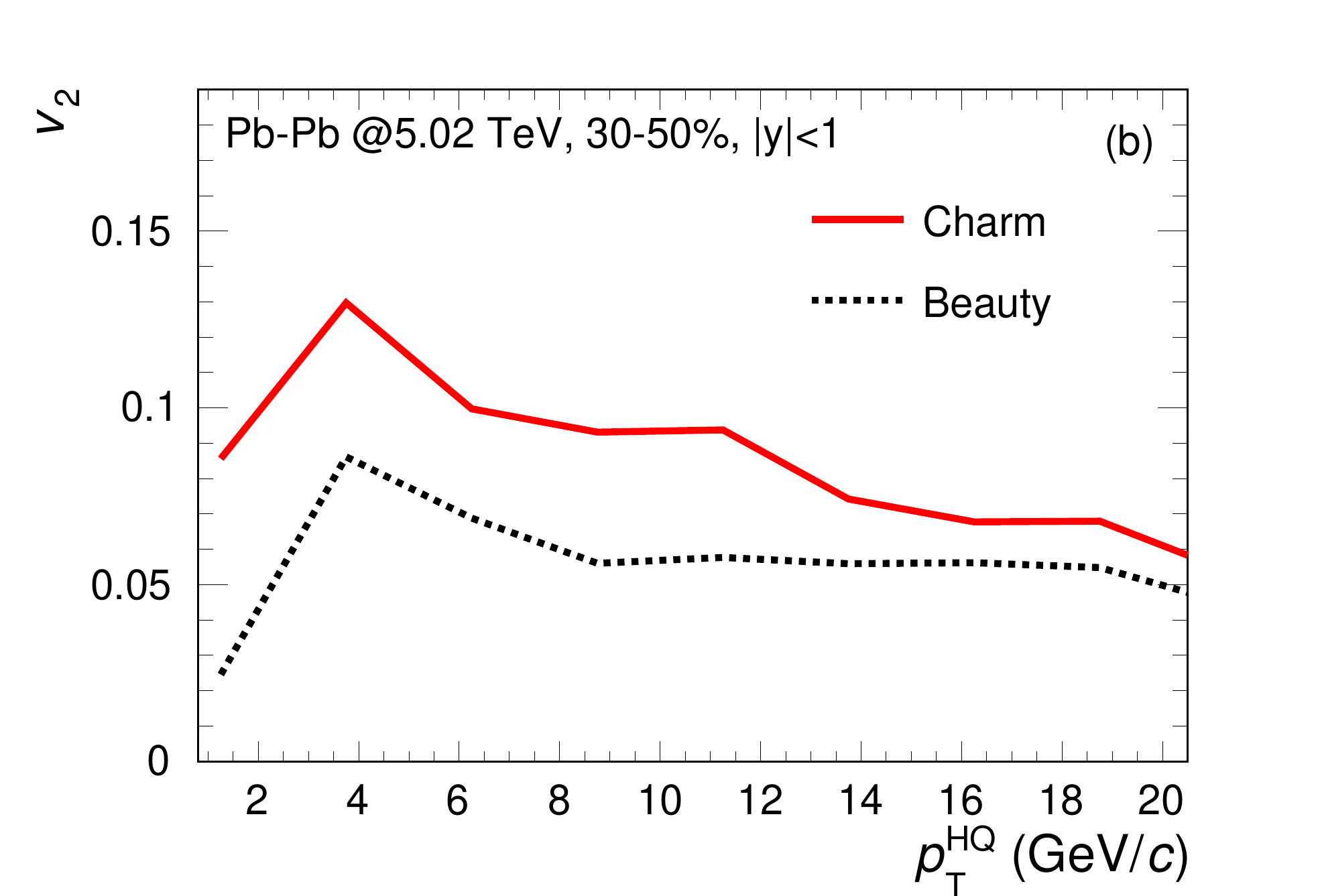}
\caption{Upper: nuclear modification factor $\raa$ for charm (solid red curve) and beauty quark (dashed black curve)
in semi-central ($30-50\%$) Pb--Pb collisions at $\snn=5.02~{\rm TeV}$;
Bottom: in analogy with the upper panel but for the azimuthal anisotropy coefficient $\vtwo$.}
\label{fig:RAAV2_HQ_5020}
\end{figure}
Figure~\ref{fig:RAAV2_HQ_5020} shows the nuclear modification modification factor $\raa$ (upper) and
the azimuthal anisotropy coefficient $\vtwo$ (bottom) of charm (solid red curve) and beauty quarks (dashed black curve)
obtained by considering both the collisional and radiative energy loss mechanisms,
in semi-central ($30-50\%$) Pb--Pb collisions at $\snn=5.02~{\rm TeV}$.
As compared to beauty, charm quark $\raa$ is more suppressed from $\pt^{\rm HQ}\sim1-2~{\rm GeV/\it{c}}$,
while it is enhanced at lower $\pt^{\rm HQ}$.
This is because the drag force (Eq.~\ref{eq:DragForce}, \ref{eq:LTECoef} and \ref{eq:Assum}) acted on the charm and beauty quarks behaves
$\frac{F_{\rm Drag}^{\rm c}}{F_{\rm Drag}^{\rm b}} \sim 0.85 \sqrt{\frac{p^{2}+m^{2}_{\rm b}}{p^{2}+m^{2}_{\rm c}}}$,
which is larger than unity below $p\sim m_{\rm b}$,
indicating that more charm quarks will be dragged toward lower $p$.
Meanwhile, the late-stage collective flow effect allows to transport the HQ from low to high $p$,
and it is more significant for the HQ with larger mass.
Thus, it increases further the difference between charm and beauty at low $p$.
As observed in Fig.~\ref{fig:DeltaE_CB_5020}, the inelastic energy loss induced by the gluon bremsstrahlung dominates at $p\gg m_{\rm b}$.
Due to the dead-cone effect, beauty quark will lose less its initial energy,
resulting in $\raa({beauty})>\raa({charm})$ in this region.
In addition to the transport coefficients,
the $\pt$-dependence of $\raa$ can also be affected by the slope of the initial momentum spectrum,
which is harder for beauty quarks, leading to a less suppression behavior of beauty quarks.
It is known that, to build a sizeable flow, the massive HQ needs frequent interactions with large coupling.
Therefore, comparing with beauty, charm quark can pick-up more $\vtwo$ from the underlying medium,
resulting in $\vtwo({charm}) > \vtwo({beauty})$ as shown clearly in the bottom panel of Fig.~\ref{fig:RAAV2_HQ_5020}.

\begin{figure}[!htbp]
\centering
\setlength{\abovecaptionskip}{-0.1mm}
\includegraphics[width=.40\textwidth]{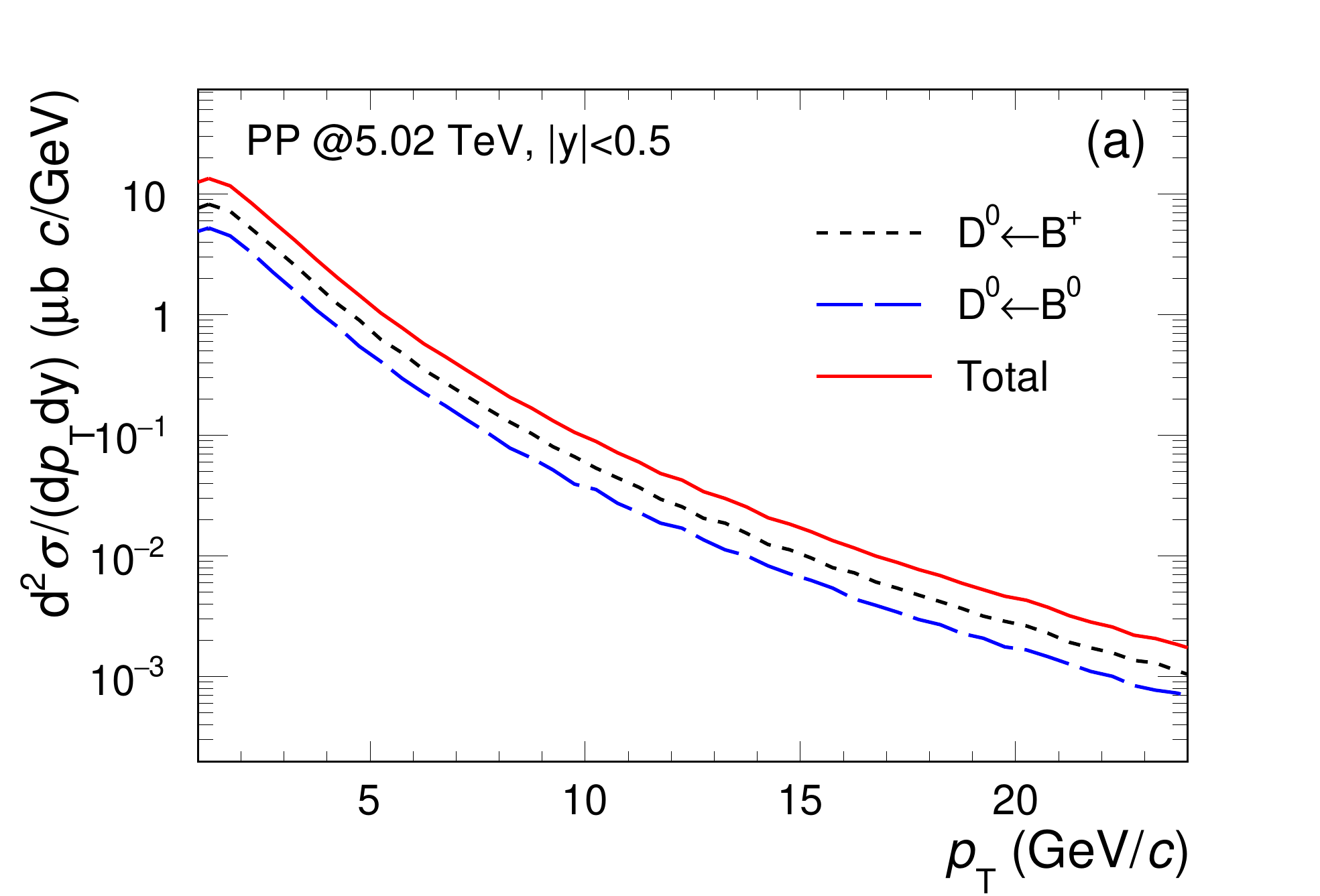}
\includegraphics[width=.40\textwidth]{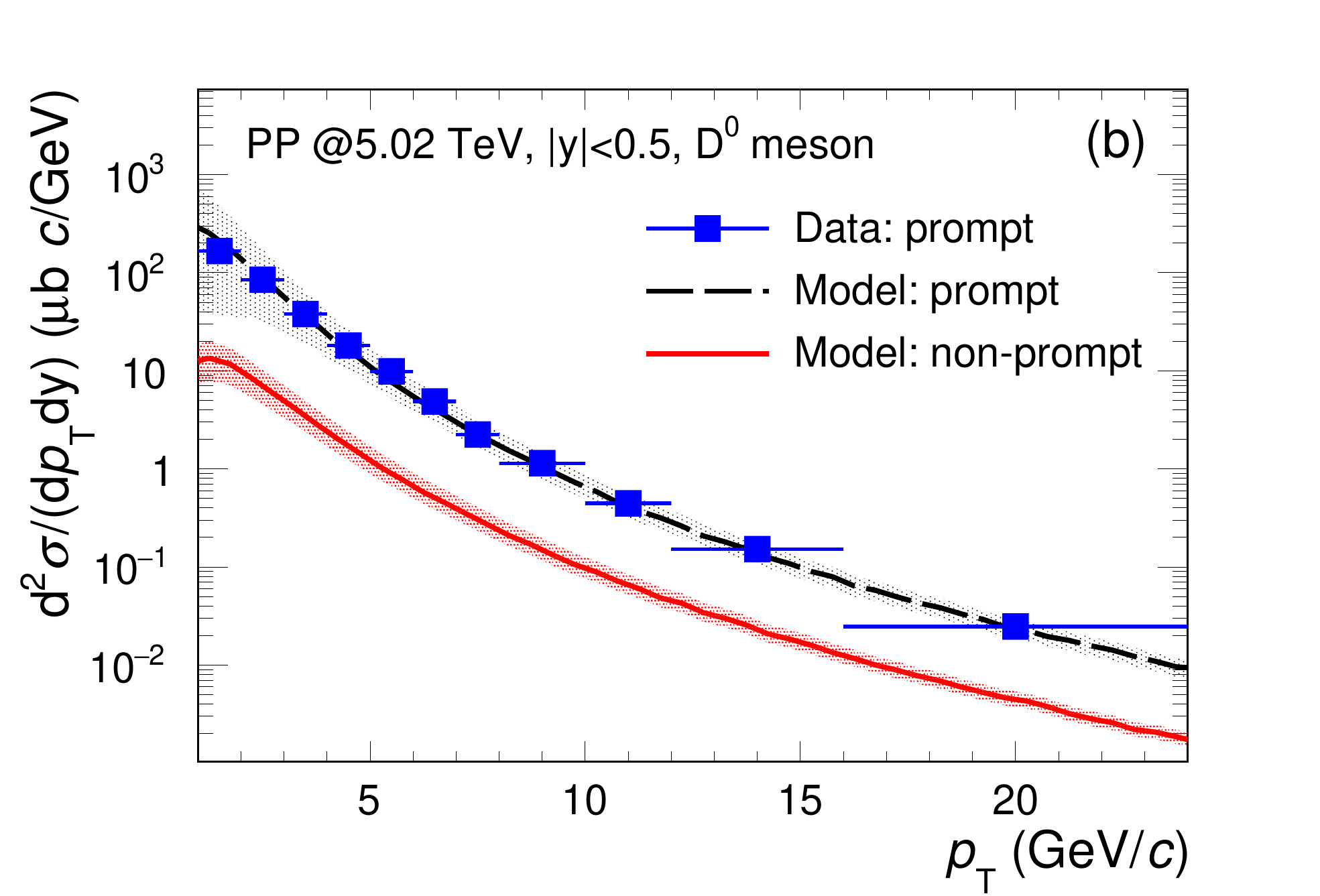}
\caption{Upper: comparison of the central values of the $\pt$-differential production cross section for
non-prompt $D^{0}$ mesons at midrapidity ($|y|<0.5$) in pp collisions at $\s=5.02~{\rm TeV}$
obtained from different decay channels: $D^{0}\leftarrow B^{+}$ (dashed black curve) and $D^{0}\leftarrow B^{0}$ (long dashed blue curve);
The combined result (solid red curve) is shown as well;
Bottom: same as above but for the comparisons between prompt and non-prompt productions
from the model and the available experimental data~\cite{DMesonPPEPJC19}.}
\label{fig:XSEC_5020}
\end{figure}
The $\pt$-differential production cross section of $D^{0}$ mesons, at midrapidity ($|y|<0.5$) in pp collisions at $\s=5.02~{\rm TeV}$,
are shown in Fig.~\ref{fig:XSEC_5020}.
In panel-a, the central values of the non-prompt component are calculated in two decay channels:
$D^{0}\leftarrow B^{+}$ (dashed black curve; $BR\approx0.876$~\cite{PDG2016})
and $D^{0}\leftarrow B^{0}$ (long dashed blue curve; $BR\approx0.555$~\cite{PDG2016}).
The combined result is presented as the red curve,
while the one including the theoretical uncertainties (red bands) are displayed in the panel-b.
The non-prompt component can be compared with the upcoming measurements at the LHC.
Meanwhile, the prompt $D^{0}$ production (black bands) is shown together with
the available measurement (blue square~\cite{DMesonPPEPJC19}).
Within the uncertainties, the experimental data is better described by the model calculation.

\begin{figure}[!htbp]
\centering
\setlength{\abovecaptionskip}{-0.1mm}
\includegraphics[width=.49\textwidth]{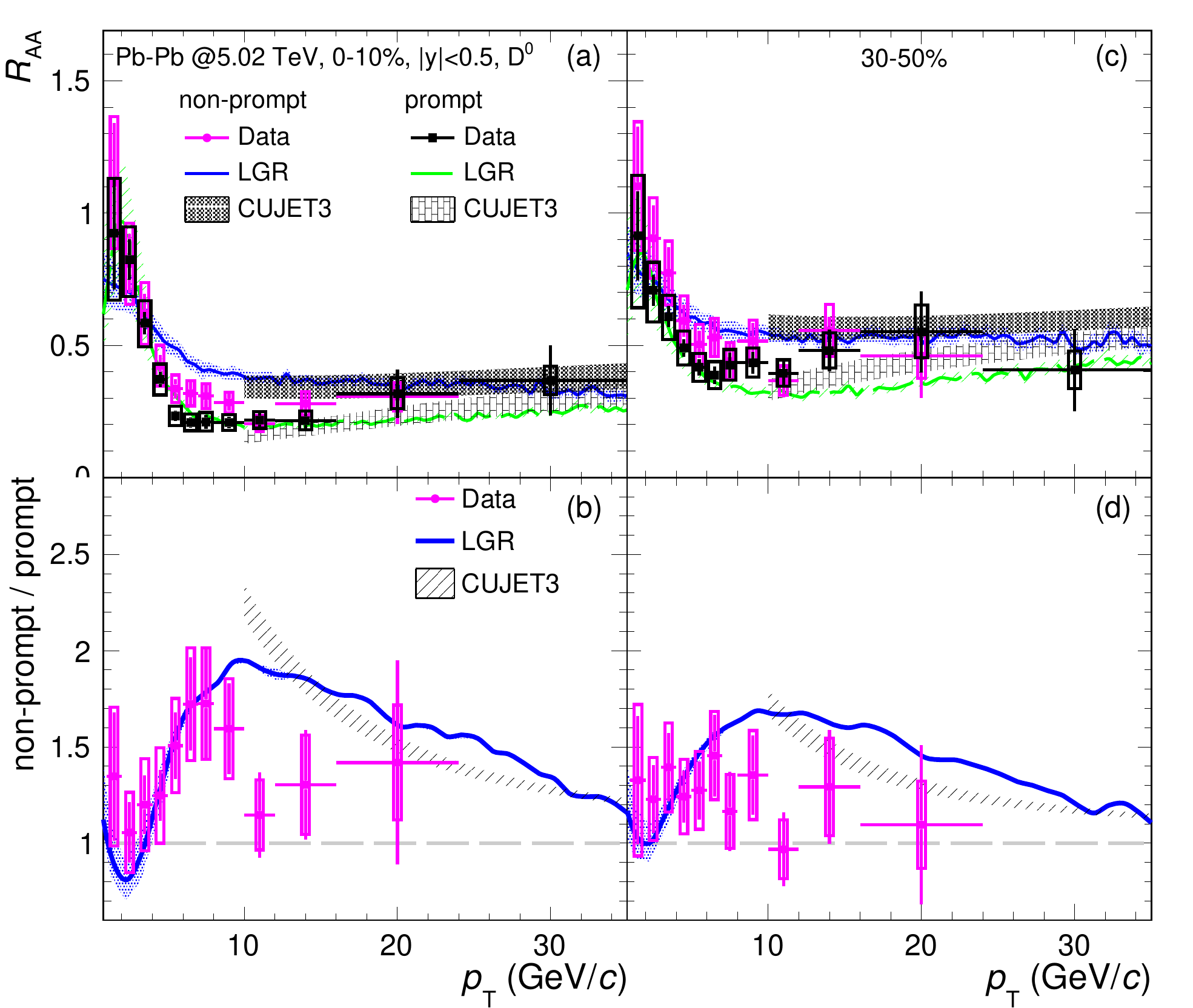}
\caption{Upper: comparison of non-prompt (solid blue curve) and prompt $D^{0}$ (dashed green curve)
$\raa$ calculations as a function of $\pt$ with the CUJET3~\cite{CUJET3JHEP16,CUJET3CPC18,CUJET3Arxiv18} (filled black regions)
and the measured values (pink circle~\cite{NonPromptDinALICE} and black square points~\cite{ALICEDesonPbPb5020RAA}),
in the central $0-10\%$ (a) and semi-central $30-50\%$ (c)
Pb--Pb collisions at $\snn=5.02~{\rm TeV}$;
Bottom: further comparison of $D^{0}$ $\raa$ between non-prompt and prompt components.
See legend and text for details.}
\label{fig:RAA_5020}
\end{figure}
\begin{figure}[!htbp]
\centering
\setlength{\abovecaptionskip}{-0.1mm}
\includegraphics[width=.38\textwidth]{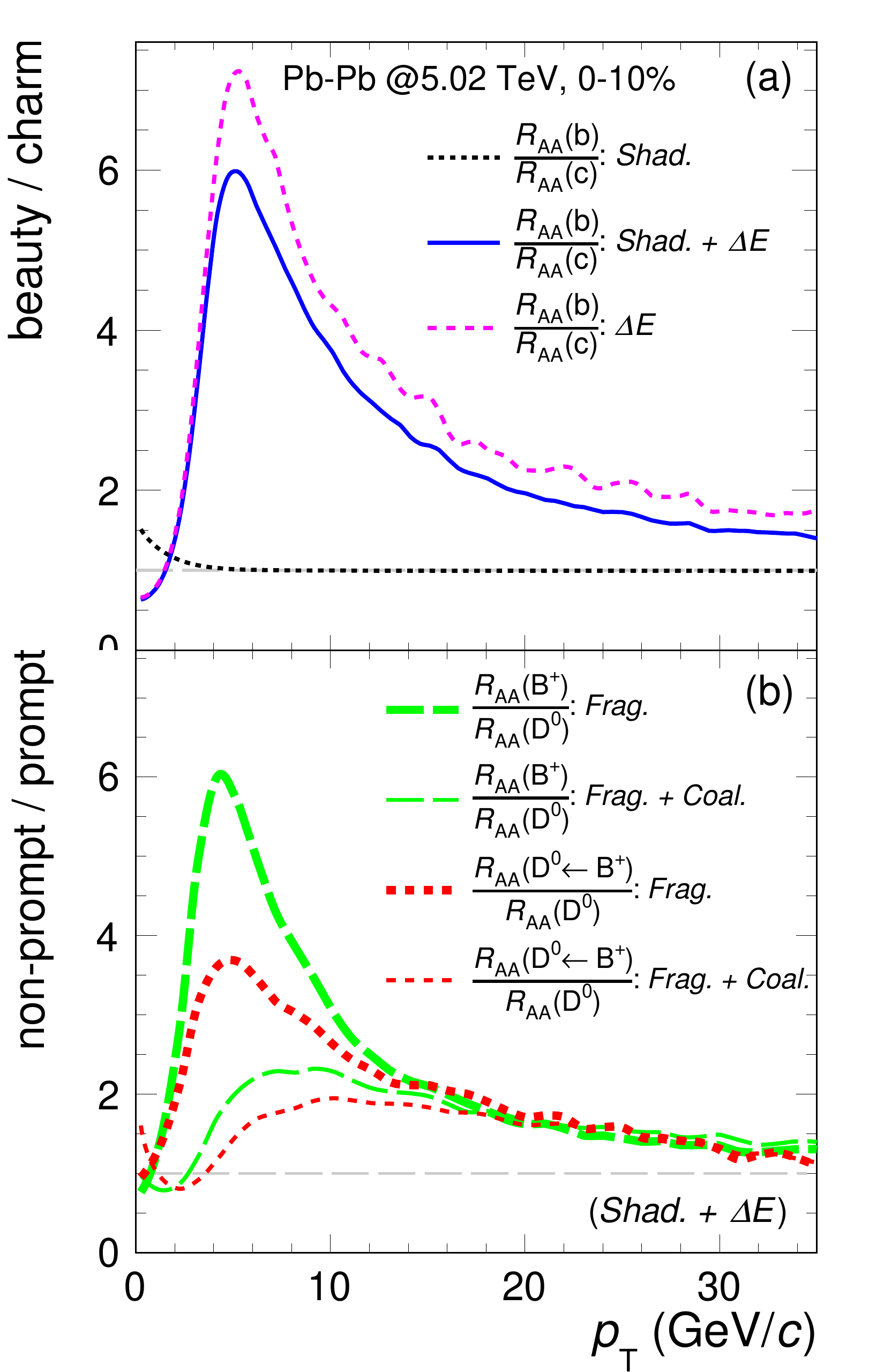}
\caption{Upper: the double ratio of the $\raa$ between beauty and charm with LGR model
in central ($0-10\%$) Pb--Pb collisions at $\snn=5.02~{\rm TeV}$.
The curves in different styles indicate the results
based on different effects: nuclear shadowing (``$Shad.$''),
in-medium energy loss together with collective expansion (``$\Delta{E}$'');
Bottom: same as above but for the heavy-flavor productions
via fragmentation (``$Frag.$'') and heavy-light coalescence (``$Coal.$'').
See legend for details.}
\label{fig:RAADoubleRatio}
\end{figure}

The panel-a in Fig.~\ref{fig:RAA_5020} shows the $\pt$-differential $\raa$ of non-prompt and prompt $D^{0}$ mesons
as a function of $\pt$ in central ($0-10\%$) Pb--Pb collisions at $\snn=5.02~{\rm TeV}$.
For comparison, the CUJET3 predictions~\cite{CUJET3JHEP16,CUJET3CPC18,CUJET3Arxiv18}
are presented as the filled black regions at $\pt>10~{\rm GeV/\it{c}}$.
By considering the full theoretical uncertainties, the two model calculations are consistent for both the non-prompt and prompt results
within the overlap $\pt$ region.
The measured $\pt$-dependence of the non-prompt $D^{0}$ $\raa$ (pink circle~\cite{NonPromptDinALICE})
is slightly overestimated by the LGR model (solid blue curve) in the range $6\lesssim \pt \lesssim 12~{\rm GeV/\it{c}}$,
while a better description observed at higher $\pt$.
Concerning the prompt $D^{0}$ $\raa$, within the experimental uncertainties,
the data (black square~\cite{ALICEDesonPbPb5020RAA}) is well reproduced by the LGR calculation (dashed green curve) in the whole $\pt$ region.
To quantify the difference between the non-prompt and prompt $D^{0}$ $\raa$, as displayed in panel-b,
a double ratio between $\raa({non-prompt})$ and $\raa({prompt})$,
is obtained and shown with the experimental data (pink circle points~\cite{NonPromptDinALICE})
and the model calculations with both the LGR (solid blue curve) and CUJET3 approaches (filled black region).
As observed in data, the double ratio is systematically larger than unity, depending on $\pt$,
reflecting a less suppression behavior for $D^{0}$ from $B$-hadron decays, as indicated in Fig.~\ref{fig:RAAV2_HQ_5020} (panel-a).
Moreover, the double ratio tends to increase in the range $2\lesssim\pt\lesssim10~{\rm GeV/\it{c}}$,
and then followed by a decreasing trend at higher $\pt$.
(See Fig.~\ref{fig:RAADoubleRatio} for further discussions).
Finally, a good agreement is found between the LGR model and the measurement at $\pt\lesssim10~{\rm GeV/\it{c}}$,
while a visible discrepancy observed around $\pt\sim12-14~{\rm GeV/\it{c}}$.
Similar conclusions can be drawn in semi-central collisions ($30-50\%$),
as shown in the panel-c and panel-d of Fig.~\ref{fig:RAA_5020}.
Note that the future measurements performed simultaneously for the non-prompt and prompt $D^{0}$ productions
are powerful to narrow down the current experimental uncertainties,
which will largely improve the data-to-model comparisons.
There are also many other independent heavy-flavor studies at high $\pt$~\cite{DjordjevicPRC15, AdsCFTPRD15, ScetPRL15, ScetPRC16, ScetJHEP17}
that deserve comparison in our future work.

To explore the underlying mechanisms of the double ratio in data (bottom panels in Fig.~\ref{fig:RAA_5020}),
in particular its bumpy structure as obsered in the measured $\pt$ range,
it is essential to quantify various effects,
such as the initial nuclear shadowing, the subsequent in-medium energy loss and collective flow,
the late-stage hadronization and the further decay, on charm and beauty quarks,
as well as on their heavy-flavor productions.
Figure~\ref{fig:RAADoubleRatio} shows the above effects on this double ratio at both parton (panel-a) and hadron levels (panel-b),
using the LGR model for central Pb--Pb collisions at $\snn=5.02~{\rm TeV}$.
For the initial produced charm and beauty quarks,
$\frac{R_{\rm AA}(b)}{R_{\rm AA}(c)}$ equals to unity without considering the nuclear (anti-)shadowing effect (dashed gray line),
which will enhance the double ratio at low $\pt$, but slightly contribute at high $\pt$ (dotted black curve).
It is caused by the fact that Bjorken-$x$ is smaller for charm quarks below $\pt\sim m_{\rm Q}$, leading to stronger suppression of $\raa(c)$,
while it is similar for both charm and beauty quarks in the range $\pt \gg m_{\rm Q}$.
By including the medium effect, e.g. the medium-induced energy loss and the hydrodynamic expansion,
the comparison between $R_{\rm AA}(b)$ and $R_{\rm AA}(c)$ is discussed in Fig.~\ref{fig:RAAV2_HQ_5020},
and the relevant double ratio is found to be largely enhanced from unity,
reaching a maximum value $\sim6$ at $\pt\sim5~{\rm GeV/\it{c}}$ (solid blue curve)
and following by a decreasing trend at higher $\pt$.
Note that, in the absence of nuclear shadowing,
the above maximum is further increased up to $\sim7$ at a similar $\pt$ value (dashed pink curve).
After the hadronization, the double ratio, including the nuclear shadowing and in-medium energy loss effects,
can also be calculated by using the resulting heavy-flavor productions,
as shown in the panel-b of Fig.~\ref{fig:RAADoubleRatio}.
We can see that, by considering alone the fragmentation,
$\frac{R_{\rm AA}(B^{+})}{R_{\rm AA}(D^{0})}$ (thick long dashed green curve) is close to the one
obtained by slightly shifting $\frac{R_{\rm AA}(b)}{R_{\rm AA}(c)}$ (solid blue curve in panel-a) toward low $\pt$.
After including the missing coalescence effect,
the peak structure of $\frac{R_{\rm AA}(B^{+})}{R_{\rm AA}(D^{0})}$ is largely broadened
with its maximum value $\sim2.5$ at $\pt\sim6-10~{\rm GeV/\it{c}}$;
however, the tail in the range $\pt\gtrsim12-14~{\rm GeV/\it{c}}$ is almost not affected.
When comparing $\frac{R_{\rm AA}(B^{+})}{R_{\rm AA}(D^{0})}$ (thick long dashed green curve)
with $\frac{R_{\rm AA}(D^{0}\leftarrow B^{+})}{R_{\rm AA}(D^{0})}$ (thick dashed red curve),
it is found that the decay kinematics of $B^{+}$
allows to suppress the peak behavior at a certain amount at a similar $\pt$.
The result for $\frac{R_{\rm AA}(D^{0}\leftarrow B^{+})}{R_{\rm AA}(D^{0})}$,
including both fragmentation and coalescence,
displays a minimum around $\pt\sim2~{\rm GeV/\it{c}}$ (thin dashed red curve),
which is mainly contributed by the coalescence
as compared to the one without including this effect (thick dashed red curve).
All the conclusions obtained above can also be drawn with $B^{0}$ and $D^{0}\leftarrow B^{0}$.

\begin{figure}[!htbp]
\centering
\setlength{\abovecaptionskip}{-0.1mm}
\includegraphics[width=.40\textwidth]{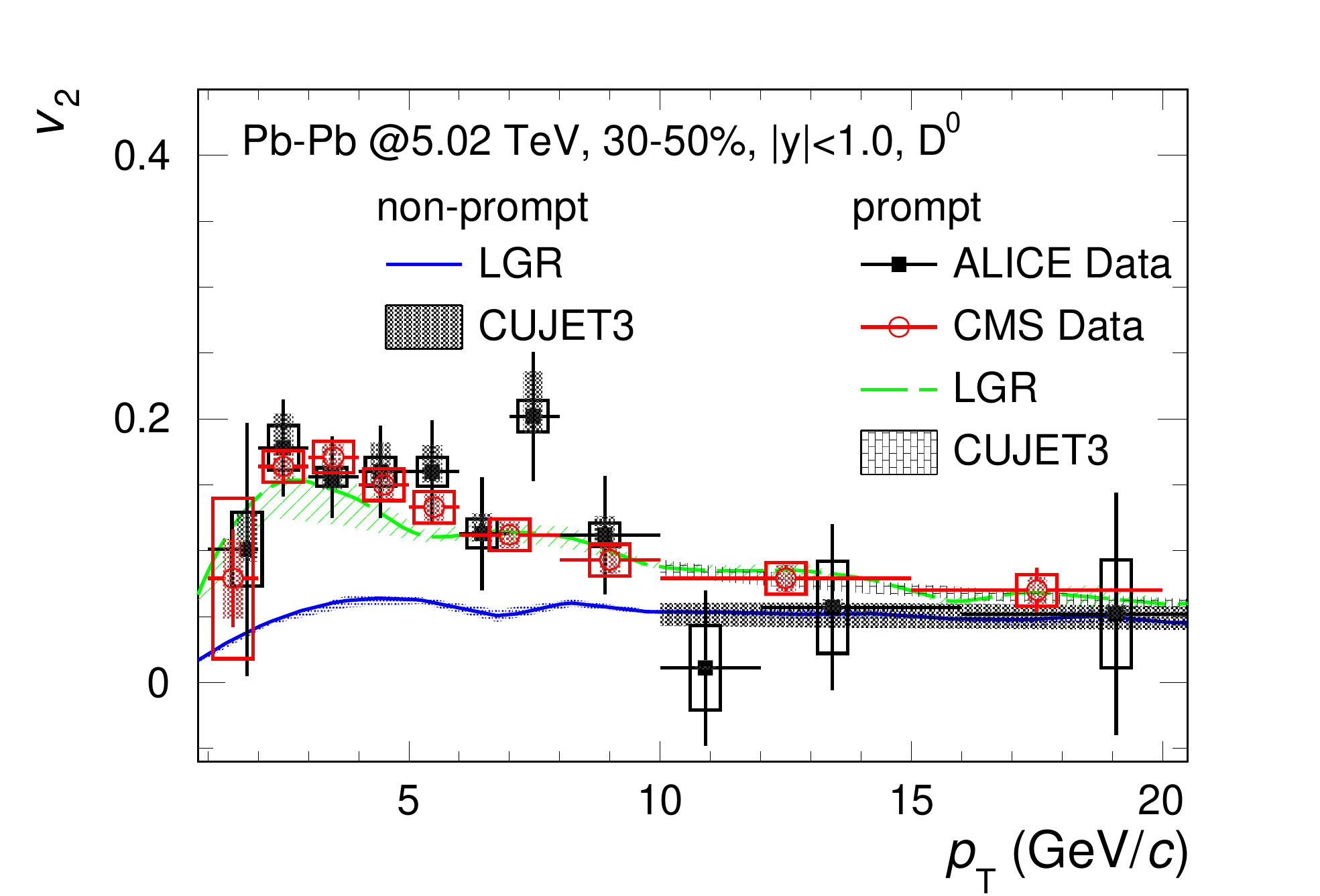}
\caption{Same as Fig.~\ref{fig:RAA_5020} (panel-c) but for $\vtwo$.
The measured $\vtwo$ is taken from ALICE (black square~\cite{ALICEDesonPbPb5020V2})
and CMS measurements (open red circle~\cite{CMSd05020V2}).}
\label{fig:V2_5020}
\end{figure}
Following Fig.~\ref{fig:RAA_5020}, Fig.~\ref{fig:V2_5020} shows also comparisons to CUJET3 predictions,
as well as to the available measurements,
but for the azimuthal anisotropy coefficient $\vtwo$ up to $\pt=20~{\rm GeV/\it{c}}$.
It is clearly found that $\vtwo({prompt}) > \vtwo(non-$ $prompt)$ > 0, espectially at $\pt\lesssim10~{\rm GeV/\it{c}}$,
revealing a significant interaction of charm and beauty quarks with the medium in this region.
Meanwhile, it exhibits a weak $\pt$-dependence at higher $\pt$,
$\vtwo({prompt}) \gtrsim \vtwo({non-prompt}) \sim 0.05$, as observed with CUJET3 approach.
During the heavy-light coalescence,
the azimuthal anisotropy of a given HQ will be increased by a certain amount,
by absorbing a coalescence partner, which consists of the medium partons.
Therefore, as expected, $\vtwo({non-prompt})$ and $\vtwo({prompt})$
are slightly larger than $\vtwo({beauty})$ and $\vtwo({charm})$, respectively.
This behavior is checked and confirmed by comparing with Fig.~\ref{fig:RAAV2_HQ_5020}.
Furthermore, the above behavior indicates that heavy-flavor meson is a genuine probe of bare heavy quark suppression.

\begin{figure}[!htbp]
\centering
\setlength{\abovecaptionskip}{-0.1mm}
\includegraphics[width=.38\textwidth]{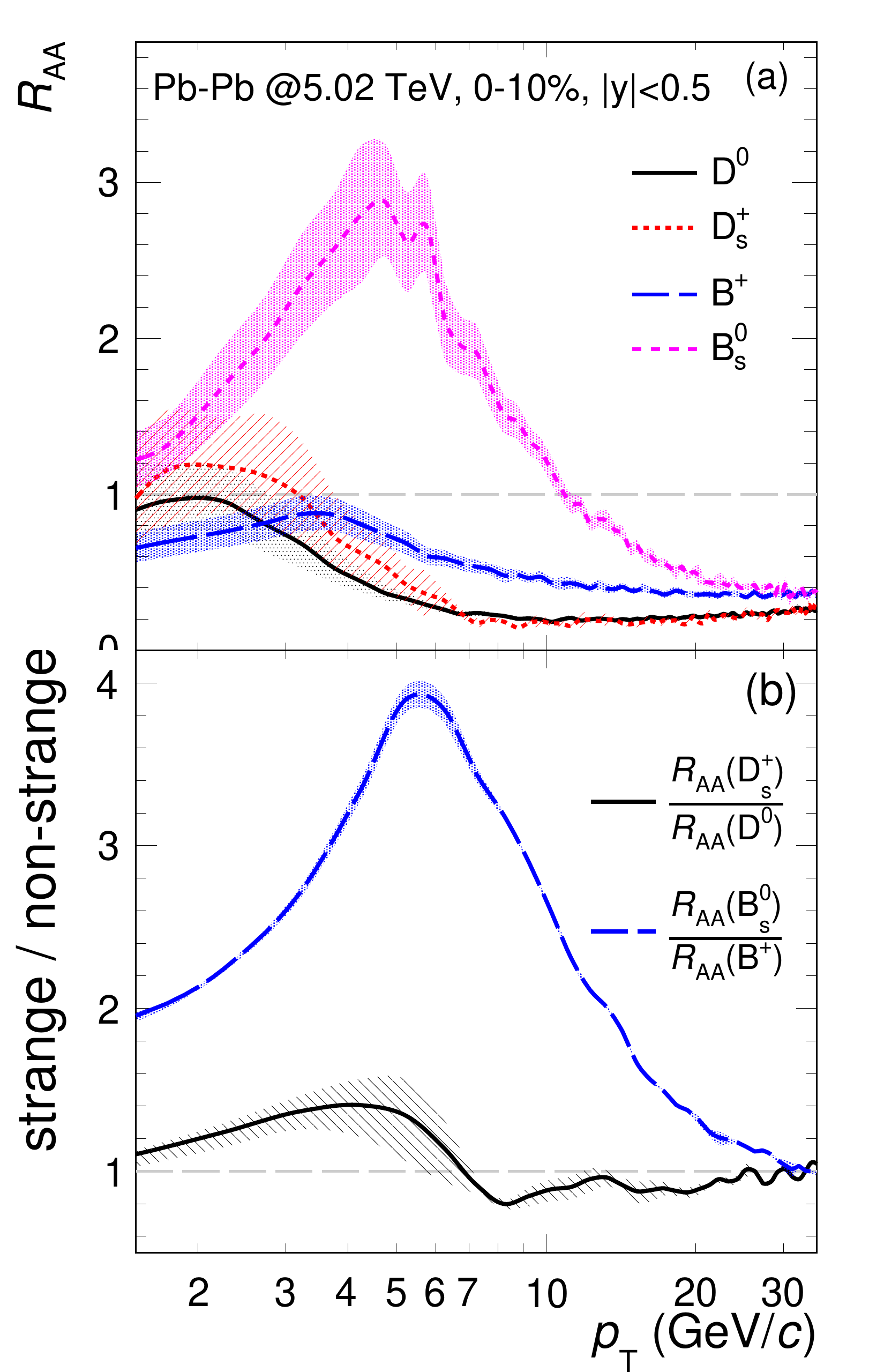}
\caption{Upper: comparison of the central predictions (curves) together with full undertainties (shadowed regions)
for $\raa$, of $D^{0}$ (solid black curve), $D^{+}_{s}$ (dotted red curve), $B^{+}$ (long dashed blue curve) and $B^{0}_{s}$ (dashed pink curve)
mesons in central ($0-10\%$) Pb--Pb collisions at $\snn=5.02~{\rm TeV}$;
Bottom: the double ratio between the strange and non-strange $\raa$ for charm (solid black curve) and beauty mesons (long dashed blue curve),
which is obtained from the upper panel.}
\label{fig:NonSvsS_5020}
\end{figure}
The final part of this section is devoted to the comparison of $\raa$ between strange and non-strange heavy-flavor mesons.
In the panel-a of Fig.~\ref{fig:NonSvsS_5020}, we present the calculations of $D^{0}({c\bar{u}})$, $D^{+}_{s}(c\bar{s})$,
$B^{+}(\bar{b}u)$ and $B^{0}_{s}(\bar{b}s)$ mesons at $|y|<0.5$ in the most central ($0-10\%$) Pb--Pb collisions $\snn=5.02~{\rm TeV}$,
as well as the double ratios between strange and non-strange, i.e. $\raa(strange)/\raa(non-strange)$,
shown in the panel-b (see legend for details).
It is known~\cite{CTGUHybrid2} that the heavy-light coalescence effect on $\raa$,
is more pronounced for $D^{+}_{s}$ as compared to $D^{0}$ mesons, in particular at intermediate $\pt$,
resulting in $\raa(D^{+}_{s})/\raa(D^{0})>1$, as displayed in panel-b.
This behavior tends to decrease at high $\pt$, where the fragmentation effect becomes significant.
The same conclusion can be drawn for beauty mesons.
By comparing $\raa(D^{+}_{s})$ with $\raa(B^{0}_{s})$,
it is found that the strangness enhancement effect is more significant for strange beauty mesons.
This is expected since the coalescence probability (Eq.~\ref{eq:InteWigner2}) for the combination $Q\bar{s}$
is larger for $Q=b$ when comparing with $Q=c$ within the considered $\pt$ region.
The observation is consistent with the $B$ ($|y|<2.4$~\cite{CMSbs5020PLB19}) and $C$-meson ($|y|<1.0$~\cite{CMSdsMesonPAS19})
measurements ($0-100\%$) with the LHC-CMS detector.

\section{Summary}\label{sec:Summary}
In summary we have studied the QGP transport properties probed by both charm and beauty quarks
by employing a recently developed framework LGR (Langevin-transport with Gluon Radiation),
which was optimized according to a global $\chi^2$ analysis in nailing down the temperature dependence
of the transport coefficients.
In particular, we've calculated $\raa$ and $\vtwo$ of charm and beauty quarks
together with their heavy-flavor productions,
and then compared with the well-known model CUJET3,
as well as the latest LHC data on open charm and beauty mesons in Pb--Pb collisions at $\snn=5.02~{\rm TeV}$.

It is found that, at low (high) momentum $p\lesssim 2m_{\rm Q}$ ($p\gg m_{\rm Q}$),
the in-medium energy loss effect is dominated by the collisional (radiative) component,
which is stronger for charm due to the mass effect $m_{\rm c}<m_{\rm b}$,
resulting in a smaller $\raa$ as compared to beauty quarks in this region.
Besides, the mass effect also allows charm quarks to pick-up more azimuthal anisotropy
during the HQ-medium interactions, leading to a systematically larger $\vtwo$ within the $\pt$ range of interest.
When considering the hadronizations including both fragmentation and coalescence mechanisms,
the above behaviors found at parton level are well inherited by the $D^{0}$ mesons from
the prompt ($D^{0}\leftarrow c$) and non-prompt processes ($D^{0}\leftarrow B \leftarrow b$).
The LGR predictions are consistent with the CUJET3 approach within the overlap $\pt$ region.
Moreover, within the experimental uncertainties,
the measured $\pt$-dependent $\raa$ and $\vtwo$ for the prompt $D^{0}$ mesons,
can be simultaneously described by LGR in central ($0-10\%$) and semi-central ($30-50\%$) Pb--Pb collisions at $\snn=5.02~{\rm TeV}$.
However, the non-prompt $\raa$ data is slightly overestimated,
especially in the range $5\lesssim\pt\lesssim12~{\rm GeV/\it{c}}$ in central collisions,
indicating an insufficient energy loss for beauty quarks in the current LGR model.
The effects such as nuclear shadowing, medium-related ones (e.g. in-medium energy loss and collective motion),
hadronization (i.e. fragmentation and coalescence) and further decay are responsible for understanding
the bumpy structure as observed in the double ratio $\raa(non-prompt)/\raa(prompt)$
around $\pt=8~{\rm GeV/\it{c}}$ in the most central region ($0-10\%$).
By comparing the strange and non-strange $D$ and $B$ mesons,
the heavy-light coalescence mechanism is more significant for strange beauty mesons.

Finally, we end with a discussion on further extending the present analysis for a followup study.
An important step further is to improve the estimation
$2\pi TD_{s}(beauty) \approx 0.85\times 2\pi TD_{s}(charm)$
which has been adopted in this work.
See the statement around Eq.~\ref{eq:Assum} for details.
The more realistic relation between $2\pi TD_{s}(beauty)$ and $2\pi TD_{s}(charm)$,
may help to reduce the discrepancy with non-prompt $\raa$ data, as mentioned above.
We plan to employ a weakly-coupled approach~\cite{POWLANGEPJC11,PBGPRC08}
and to further extended to include the missing inelastic scatterings.
After that, we will update the current results by including also the data collected at RHIC energy.
\begin{acknowledgements}
The authors thank Andrea Beraudo, Miklos Gyulassy, Jinfeng Liao, Xinye Peng, Shuzhe Shi, Hongxi Xing, Xiaoming Zhang and
Pengfei Zhuang for their kind help and useful discussions.
S.~L. is supported by the Hubei Provincial Natural Science Foundation under Grant No.2020CFB163,
the National Science Foundation of China (NSFC) under Grant Nos.12005114, 11847014 and 11875178,
and the Key Laboratory of Quark and Lepton Physics Contracts No. QLPL2018P01.
R.~W. acknowledges the support from NSFC under Project No.11505130.
\end{acknowledgements}
%
\bibliography{Shuang_p5}
\end{document}